# Pseudo-Hermiticity protects the energy-difference conservation in the scattering

H. S. Xu and L. Jin[*]
*School of Physics, Nankai University, Tianjin 300071, China*

Symmetry plays a fundamentally important role in physics. In this work, we find a conservation law, $S^{\dagger}(H_c^{\dagger})S(H_c) = I$, which is valid for any non-Hermitian scattering center $H_c$. As a result, the reflections and transmissions of a non-Hermitian system $\{r, t\}$ and its Hermitian conjugation system $\{\bar{r}, \bar{t}\}$ satisfy the conservation law $\bar{r}^*r + \bar{t}^*t = 1$, instead of the energy conservation law that applies to incoming and outgoing waves in a Hermitian system. Consequently, the pseudo-Hermiticity of a non-Hermitian system ensures an energy-difference conservation. Furthermore, we demonstrate that the energy-difference conservation is respectively valid and invalid in two prototypical anti-$\mathcal{PT}$-symmetric systems, where the energy-difference conservation is protected by the pseudo-Hermiticity. Our findings provide profound insight into the conservation law, the pseudo-Hermiticity, and the anti-$\mathcal{PT}$-symmetry in non-Hermitian systems.

*Introduction.* Open systems exchange energy with their environment in contrast to the energy conservation in the closed systems. Therefore, the Hamiltonians of the open systems are non-Hermitian [1]. In a general case, the spectra of non-Hermitian systems are complex, the eigenstates are nonorthogonal, and the dynamics are nonunitary [2–26]. Counterintuitively, the non-Hermiticity arising from the interaction with the environment can cause many intriguing phenomena, including the coherent perfect absorption [27–40], the loss induced revival of lasing [41], the exceptional point enhanced sensing [42–50], the asymmetric mode switching [51–57], and the exotic topology of exceptional points [58–66].

The excitation intensity in the exact parity-time ($\mathcal{PT}$) symmetric phase experiences an oscillation even though their energy spectrum is entirely real under the $\mathcal{PT}$ symmetry protection [67–77]. Although non-Hermitian systems have a real spectrum, their eigenstates are not orthogonal due to non-Hermiticity, which is the origin of nonunitary dynamics. However, non-Hermitian Noether's theorem has revealed a set of conserved quantities [78–83]. Furthermore, unitary dynamics are possible in non-Hermitian systems. When excitations involve only real-valued, orthogonal eigenstates in non-Hermitian systems, time evolution is always unitary, and energy conservation is still observed [84]. In addition, energy conservation is possible before and after scattering by a non-Hermitian scattering center, with the scattering matrix being unitary, despite the dynamics throughout the scattering process being nonunitary due to the effects of non-Hermiticity [85]. Interestingly, the energy-difference relation has been demonstrated in the anti-$\mathcal{PT}$-symmetric system [86].

Here, we report a universal conservation law for every non-Hermitian scattering system. The validity of conservation is independent of the degree of non-Hermiticity and the reality of the scattering center spectrum. Then we find that the reflections and transmissions of a non-Hermitian system $\{r, t\}$ and its Hermitian conjugation system $\{\bar{r}, \bar{t}\}$ satisfy $\bar{r}^*r + \bar{t}^*t = 1$. Consequently, the conservation can be established from the symmetry protection of a specific pseudo-Hermiticity of the non-Hermitian system. Furthermore, we clarify that the energy-difference conservation in the anti-$\mathcal{PT}$-symmetric systems is protected by this specific pseudo-Hermiticity. Our findings are applicable in a variety of physical systems, including cold atomic gases, coupled resonators/waveguides/fibers, electrical circuits, heat diffusion systems, cavity magnonics, plasmonics, and beyond.

*Conservation law in the scattering.* Symmetries play an important role in the non-Hermitian systems. For example, the most investigated parity-time ($\mathcal{PT}$) symmetry $(\mathcal{PT})H(\mathcal{PT})^{-1} = H$ and anti-parity-time (anti-$\mathcal{PT}$) symmetry $(\mathcal{PT})H(\mathcal{PT})^{-1} = -H$ impose unique properties and stimulate novel applications in physics. The $\mathcal{PT}$ symmetry is able to create an entirely real spectrum. The $\mathcal{PT}$ symmetry breaks as a phase transition at the exceptional point. The $\mathcal{PT}$ symmetry also ensures reciprocal transmission [87–91]. The anti-$\mathcal{PT}$-symmetry is able to create an entirely imaginary spectrum. The anti-$\mathcal{PT}$-symmetry counterintuitively ensures the energy-difference conserving dynamics [86,92]; by contrast, the energy is conserved before and after a nonunitary scattering process and the scattering matrix is unitary [85,93].

We provide a proof of conservation in the non-Hermitian scattering systems from the viewpoint of the biorthogonal quantum mechanics [83,94]. We consider a general time-evolution problem with two initial states $\varphi(0)$ and $\phi(0)$. The time-evolution states of $\varphi(0)$ in a non-Hermitian system $H$ and $\phi(0)$ in Hermitian conjugate system $H^{\dagger}$ yield $\varphi(t) =$ $e^{-iHt}\varphi(0)$ and $\phi(t) = e^{-iH^\dagger t}\phi(0)$, respectively. Notably, their overlap is constant $\langle\phi(t)|\varphi(t)\rangle = \langle\phi(0)|\varphi(0)\rangle$. We emphasize that the results are independent of the chosen initial states and the chosen non-Hermitian systems. If we set the initial state $\varphi(0) = \phi(0)$ and the Hermitian system $H = H^\dagger$, the relation $\langle\phi(t)|\varphi(t)\rangle = 1$ reduces to the energy conservation. We apply this generalized conservation to scattering problem. The scattering solves the problems including the light propagation in the integrated photonics and the quantum transport in mesoscopic quantum systems.

---

*jinliang@nankai.edu.cn

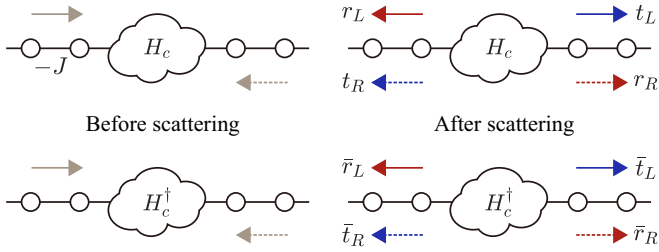

FIG. 1. Schematic of the scattering coefficients in a non-Hermitian scattering center $H_c$ and its Hermitian conjugation counterpart $H_c^\dagger$. The wave after scattering center having the opposite propagation direction is the reflection. The wave after scattering center having the same propagation direction is the transmission.

Figure 1 illustrates a two-port scattering system consisting of a scattering center and two semi-infinite uniform chains serving as the input and output ports. The scattering center Hamiltonian is $H_c$ and the port has a uniform coupling $-J$. The ports are connected to the scattering center at the coupling strength $-J$. In this sense, the scattering properties are not affected by the connections, and the reflection and transmission fully characterize the scattering properties of the scattering center $H_c$. The scattering properties are obtained from the steady-state solutions of the Schrödinger equations for the two-port scattering system. The steady state is a superposition of plane waves with opposite propagation directions.

We consider the scattering problem in the scattering center $H_c$ and its Hermitian conjugate $H_c^\dagger$. The steady state of a scattering system is a superposition of counter-propagating plane waves with opposite momentum. For a wave vector $k \in (0, \pi)$ input in the left (right) port of the scattering center $H_c$, the wave function in the left port is $e^{ikj} + r_L e^{-ikj}$ ($t_R e^{-ikj}$), and the wave function in the right port is $t_L e^{ikj}$ ($e^{-ikj} + r_R e^{ikj}$). The reflection coefficient is $r_{L(R)}$, and the transmission coefficient is $t_{L(R)}$. In this case, the incoming wave before scattering is $e^{ikj}$ ($e^{-ikj}$) in the left (right) port, while the outgoing waves after scattering are $r_L e^{-ikj}$ ($t_R e^{-ikj}$) in the left port and $t_L e^{ikj}$ ($r_R e^{ikj}$) in the right port. For a wave vector $k$ input in the left (right) port of the scattering center $H_c^\dagger$, the wave function in the left port is $e^{ikj} + \bar{r}_L e^{-ikj}$ ($\bar{t}_R e^{-ikj}$), and the wave function in the right port is $\bar{t}_L e^{ikj}$ ($e^{-ikj} + \bar{r}_R e^{ikj}$). The reflection coefficient is $\bar{r}_{L(R)}$, and the transmission coefficient is $\bar{t}_{L(R)}$. In this case, the incoming wave before scattering is $e^{ikj}$ ($e^{-ikj}$) in the left (right) port, while the outgoing waves after scattering are $\bar{r}_L e^{-ikj}$ ($\bar{t}_R e^{-ikj}$) in the left port and $\bar{t}_L e^{ikj}$ ($\bar{r}_R e^{ikj}$) in the right port.

The wave function overlap after scattering equals the wave function overlap before scattering. When both incoming waves of $H_c$ and $H_c^\dagger$ before scattering are in the left (right) port, their overlap is unity. Therefore, the overlap between the outgoing waves of $H_c$ and $H_c^\dagger$ after scattering is also unity, yielding the relation

$$\bar{r}_{L(R)}^* r_{L(R)} + \bar{t}_{L(R)}^* t_{L(R)} = 1. \quad (1)$$

When the incoming wave of $H_c$ before scattering is in the left (right) port and $H_c^\dagger$ is in the right (left) port, their wave function overlap is zero. Therefore, the overlap between the outgoing waves of $H_c$ and $H_c^\dagger$ after scattering is also zero, yielding the relation

$$\bar{t}_{R(L)}^* r_{L(R)} + \bar{r}_{R(L)}^* t_{L(R)} = 0. \quad (2)$$

The conservation relations, Eqs. (1) and (2), are outcomes of the conservation $\langle\phi(t)|\varphi(t)\rangle = \langle\phi(0)|\varphi(0)\rangle$ applied to the scattering system. Notably, these relations are valid for any non-Hermitian scattering system and Eq. (1) reduces to energy conservation for any Hermitian scattering system. The scattering matrix for the scattering center $H_c$ and $H_c^\dagger$ are defined as

$$S(H_c) = \begin{pmatrix} r_L & t_R \\ t_L & r_R \end{pmatrix}, \quad S(H_c^\dagger) = \begin{pmatrix} \bar{r}_L & \bar{t}_R \\ \bar{t}_L & \bar{r}_R \end{pmatrix}. \quad (3)$$

Therefore, Eqs. (1) and (2) yield a conservation law established from the scattering matrices in the form of

$$S^\dagger(H_c^\dagger) S(H_c) = I, \quad (4)$$

where $I$ stands for the identity matrix. The conservation law yields the fact that a Hermitian scattering system has the unitary scattering matrix. In the discussion, we prove that the conservation law is valid for arbitrary scattering system.

*Pseudo-Hermiticity symmetry protection.* The scattering properties are closely related to the symmetry of the scattering center. For example, the symmetric reflection and/or symmetric transmission can result from specific discrete symmetries [95,96]. In this work, however, we focus on the conservation rather than the symmetric dynamics. Specifically, we consider the pseudo-Hermitian links $H_c$ and $H_c^\dagger$, which satisfy

$$q H_c^\dagger q^{-1} = H_c, \quad (5)$$

where the similar transformation $q$ satisfies $q = q^\dagger$ [91]. If the scattering center $H_c$ has a specific pseudo-Hermiticity, the scattering coefficients for $H_c$ and $H_c^\dagger$ are closely related, depending on the formation of the similar transformation $q$. To proceed with our discussion, we assume that the two ports are connected to the $m$ and $n$ sites of the scattering center $H_c$. If the elements of the similar transformation satisfy $q_{mj} = q_{jm} = \delta_{mj}$ and $q_{nj} = q_{jn} = \pm\delta_{jn}$, where $\delta_{mj}$ and $\delta_{jn}$ are the Kronecker delta function, the pseudo-Hermiticity establishes the following relations:

$$\bar{r}_L = r_L, \bar{t}_L = \pm t_L, \quad \bar{r}_R = r_R, \bar{t}_R = \pm t_R. \quad (6)$$

Then, we obtain the relations between the incoming photon flux and the outgoing photon flux. The total flux of the wave injection is unchanged before and after scattering ($|r_{L(R)}|^2 + |t_{L(R)}|^2 = 1$) for $q_{mm}q_{nn} = +1$, and the reflected photon flux minus the transmitted photon flux equals to the input photon flux ($|r_{L(R)}|^2 - |t_{L(R)}|^2 = 1$) for $q_{mm}q_{nn} = -1$. In the scattering, the photons in different ports always have the same energy. Therefore, these relations also mean the energy conservation in the former case and the energy-difference conservation in the later case.

We briefly discuss $\mathcal{PT}$ symmetry, anti-$\mathcal{PT}$ symmetry, and the pseudo-Hermiticity in non-Hermitian systems. In many cases, the non-Hermitian systems under consideration also have the reciprocity $H = H^T$. In such cases, the definition of $\mathcal{PT}$ symmetry coincides with that of pseudo-Hermiticity, and the definition of anti-$\mathcal{PT}$ symmetry coincides with that of pseudo-anti-Hermiticity, which contradicts the pseudo-Hermiticity. However, the requirement $q_{mn} = q_{nm} = 0$ for the conservations $|r_{L(R)}|^2 \pm |t_{L(R)}|^2 = 1$ contradicts the required $q_{mn}q_{nm} = \pm 1$ for the $\mathcal{P}$ operator in the $\mathcal{PT}$ symmetry and anti-$\mathcal{PT}$ symmetry. Fortunately, the non-Hermitian systems may have multiple symmetries that function simultaneously, with each symmetry independently playing a unique role of symmetry protection. The energy-difference conservation holds when specific pseudo-Hermiticity protection is found. In practice, the $\mathcal{PT}$-symmetric systems typically do not hold the energy conservation or the energy-difference conservation because $q_{mn}q_{nm} = \pm 1$ is necessary to switch the balanced gain and loss for a $\mathcal{PT}$ symmetry. However, the anti-$\mathcal{PT}$-symmetric systems mostly satisfy this requirement because $q_{mn}q_{nm} = \pm 1$ is not necessary for a pseudo-Hermiticity. These arguments explain why energy conservation and energy-difference conservation are typically observed in anti-$\mathcal{PT}$-symmetric systems, but not in $\mathcal{PT}$-symmetric scattering systems.

*Anti-$\mathcal{PT}$-symmetric systems.* The prototypical anti-$\mathcal{PT}$-symmetric systems that have been mostly investigated involve imaginary couplings of $-i\gamma$, which can be realized through a dissipative mediator in various experimental platforms, including the cold atom gases [97–99], coupled waveguides [100–108], microcavities [109–116], synthetic spectral dimension [117], optical fibers [118,119], electrical circuits [86,120–123], heat diffusion systems [124], cavity magnonics [125–129], and plasmonics [130]. Notably, although many anti-$\mathcal{PT}$-symmetric systems under investigation are pseudo-Hermitian, anti-$\mathcal{PT}$ symmetry and pseudo-Hermiticity are fundamentally different. We illustrate this point using two prototypical examples that have been intensively investigated:

$$H_{c,1} = \begin{pmatrix} -i\gamma + V & -i\gamma \\ -i\gamma & -i\gamma - V \end{pmatrix}, \quad H_{c,2} = \begin{pmatrix} V & -i\gamma \\ -i\gamma & -V \end{pmatrix}. \quad (7)$$

Eliminating the dissipative mediator results in the imaginary coupling of $-i\gamma$ and the common dissipation of $-i\gamma$ to realize the non-Hermitian Hamiltonian $H_{c,1}$. The common loss of $-i\gamma$ leads to an imaginary energy shift. The two eigenenergies of $H_{c,1}$ are $-i\gamma \pm \sqrt{V^2 - \gamma^2}$, and the two eigenenergies of $H_{c,2}$ are $\pm\sqrt{V^2 - \gamma^2}$. Previous investigations have shown that the properties of $H_{c,1}$ and $H_{c,2}$ are identical, in particular, the anti-$\mathcal{PT}$-symmetry phase transition. In both systems, the exact anti-$\mathcal{PT}$-symmetric phase is when $V < \gamma$, and the broken anti-$\mathcal{PT}$-symmetric phase is when $V > \gamma$. Therefore, $H_{c,1}$ is mostly studied to uncover the anti-$\mathcal{PT}$-symmetric properties of $H_{c,2}$. However, the non-Hermitian systems $H_{c,1}$ and $H_{c,2}$ exhibit completely different properties when they are embedded in a scattering system as the scattering center to control wave transport. For example, the scattering properties of $H_{c,1} = -i\gamma(\sigma_x + \sigma_0)$ and $H_{c,2} = -i\gamma\sigma_x$ can be completely different although they are both anti-$\mathcal{PT}$-symmetric $\sigma_x H_{c,1}^* \sigma_x^{-1} = -H_{c,1}$, $\sigma_x H_{c,2}^* \sigma_x^{-1} = -H_{c,2}$, where $\sigma_{x,y,z}$ are the Pauli matrices and $\sigma_0$ is the $2 \times 2$ identity matrix. Notably, the former scattering center $H_{c,1}$ is (pseudo-)anti-Hermitian $H_{c,1}^\dagger = -H_{c,1}$, and the latter scattering center $H_{c,2}$ is pseudo-Hermitian $\sigma_z H_{c,2}^\dagger \sigma_z^{-1} = H_{c,2}$. The on-site loss in $H_{c,1}$ breaks the pseudo-Hermiticity of $H_{c,1}$, while the pseudo-Hermiticity of $H_{c,2}$ protects the energy-difference conservation. Notably, we also have the pseudo-Hermiticity $\sigma_y H_{c,2}^\dagger \sigma_y^{-1} = H_{c,2}$, which does not protect the energy-difference conservation. We emphasize that the above conclusions are protected by the symmetry of the Hamiltonian rather than the symmetry of the eigenstates. The validity of the conservation is independent of the exact/broken anti-$\mathcal{PT}$-symmetric phases and the reality of the scattering center spectrum.

We calculate the scattering coefficients of the two exemplified scattering centers and perform numerical simulations to verify our findings. Without loss of generality, we exemplify the case $V = 0$ for the conciseness. In the Supplemental Material, we show the case $V \neq 0$ [131], where the results for the scattering center in the exact anti-$\mathcal{PT}$-symmetric phase, at the anti-$\mathcal{PT}$-symmetric phase transition, and in the broken anti-$\mathcal{PT}$-symmetric phases are all shown. We point out that the scattering coefficients are not sensitive to the anti-$\mathcal{PT}$-symmetry phase transition. In the case $V = 0$, the system has the inversion symmetry, and the scattering coefficients are independent of the input direction. Consequently, both the reflection and the transmission are symmetric. Figure 2(a) schematically illustrates the scattering system $H_{c,1}$. The steady-state Schrödinger equations for the sites in the scattering center $H_{c,1}$ are $E\psi_{c,1} = -i\gamma\psi_{c,1} - i\gamma\psi_{c,2} - J\psi_{-1}$, and $E\psi_{c,2} = -i\gamma\psi_{c,2} - i\gamma\psi_{c,1} - J\psi_1$. The scattering coefficients can be straightforwardly obtained by substituting the wave functions, the dispersion relation, and the wave function continuity relations into the above Schrödinger equations

$$r = -\frac{iJ + 2\gamma \cos k}{iJ + 2\gamma e^{ik}}, \quad t = \frac{2i\gamma \sin k}{iJ + 2\gamma e^{ik}}. \quad (8)$$

Figure 2(b) schematically illustrates the Hermitian conjugation counterpart $H_{c,1}^\dagger$ of the scattering system in Fig. 2(a). Notably, $H_{c,1}^\dagger$ is obtained by substituting $\gamma$ with $-\gamma$ in $H_{c,1}$. The scattering coefficients of $H_{c,1}^\dagger$ are given by Eq. (8) after substituting $\gamma$ with $-\gamma$.

The reflection and transmission of $H_{c,1}$ are plotted in Fig. 2(c), where the reflection is indicated by the red circle, the transmission is indicated by the blue square. It is evident that the total intensity of the scattered light is less than unity, implying that the scattering center loses energy to the environment via dissipation. The difference between the reflection and transmission is the minimum at the momentum $k = \pi/2$ for the resonant input. In contrast, the reflection and transmission of $H_{c,1}^\dagger$ are plotted in Fig. 2(d). The total intensity of the scattered light is larger than unity, indicating that the scattering system extracts energy from the environment via gain in the scattering center. The difference between the reflection and transmission is the maximum at the momentum $k = \pi/2$ for the resonant input. To verify the conservation relation, the real part and the imaginary part of $\bar{r}^*r$ and $\bar{t}^*t$ are plotted in Figs. 2(e) and 2(f), respectively. The sum of the real parts is equal to unity, and the sum of the imaginary parts is equal to zero for any input momentum in the range $k \in (0, \pi)$. We perform the time-evolution simulations of a Gaussian wave packet in $H_{c,1}$ and $H_{c,1}^\dagger$ for resonant input at $k = \pi/2$, as shown in Figs. 2(g) and 2(h), respectively. The initial excitation is a normalized Gaussian wave packet given by

$$|\psi(j,0)\rangle = \Omega^{-1/2} \sum_j e^{-(j-n_0)^2/2\sigma^2} e^{ikj} |j\rangle. \quad (9)$$

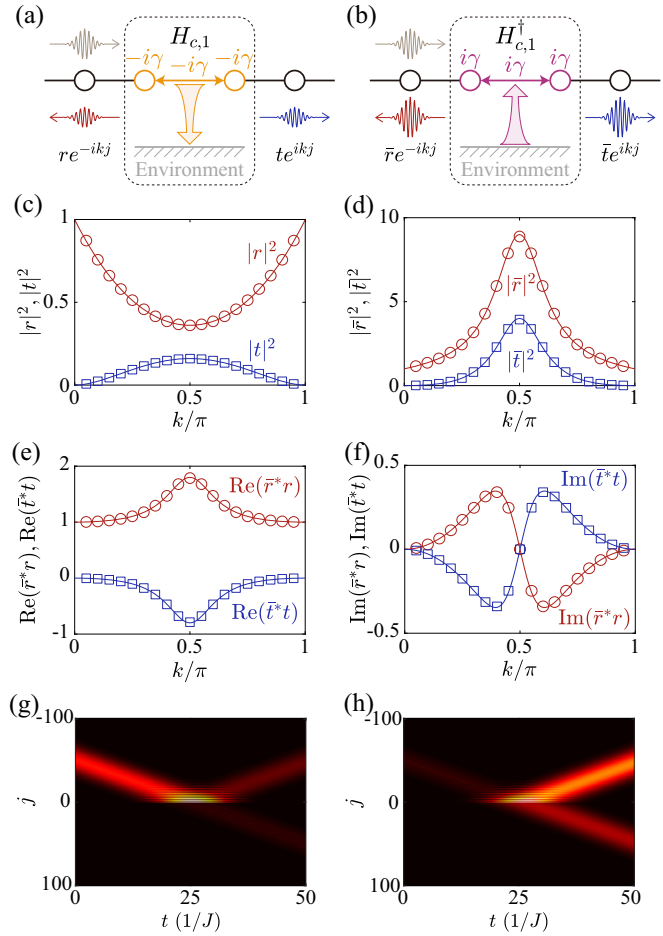

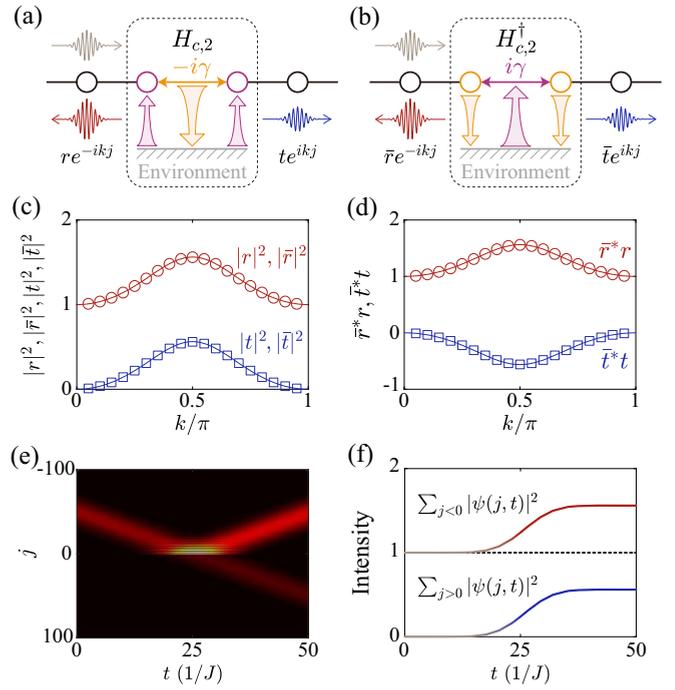

FIG. 2. (a) and (b) are schematics of the anti-$\mathcal{PT}$-symmetric scattering systems $H_{c,1}$ and $H_{c,1}^\dagger$ without the pseudo-Hermiticity protection. (c) Reflection and transmission in (a). (d) Reflection and transmission in (b). (e) The real part of $\bar{r}^*r$ and $\bar{t}^*t$. (f) The imaginary part of $\bar{r}^*r$ and $\bar{t}^*t$. The parameter is $\gamma = J/3$. The numerical simulations of Gaussian wave packet at the resonant input $k = \pi/2$ in (a) and (b) are performed in (g) and (h).

FIG. 3. (a) and (b) are schematics of the anti-$\mathcal{PT}$-symmetric scattering system $H_{c,2}$ and $H_{c,2}^\dagger$ with the pseudo-Hermiticity protection. (c) Reflection and transmission in (a) and (b). (d) $\bar{r}^*r$ and $\bar{t}^*t$. The parameter is $\gamma = J/3$. (e) Numerical simulation of Gaussian wave packet time evolution in $H_{c,2}$ for the resonant input at $k = \pi/2$. (f) The reflection and transmission in the time-evolution dynamics.

The Gaussian wave packet is centered at the site $n_0 = -50$, where $\Omega = \sqrt{\pi}\sigma$ is the normalization factor, $\sigma = 10$ controls the size, and $|j\rangle$ is the basis of port site $j$. After the incident wave packet is scattered by the scattering center, the intensity of reflected (transmitted) wave packet represents the reflection (transmission). The reflection and transmission in Fig. 2(g) are 0.36 and 0.16, respectively. The reflection and transmission in Fig. 2(h) are 8.9 and 3.9, respectively. The numerical simulation results are in good agreement with our predictions.

Figure 3(a) shows the schematic of the scattering system $H_{c,2}$. The steady-state Schrödinger equations for the scattering center sites are $E\psi_{c,1} = -i\gamma\psi_{c,2} - J\psi_{-1}$ and $E\psi_{c,2} = -i\gamma\psi_{c,1} - J\psi_1$. Similarly, the scattering coefficients can be obtained by substituting the wave functions, the dispersion relation, and the wave function continuity relations into the above Schrödinger equations

$$r = -\frac{J^2 + \gamma^2}{J^2 + \gamma^2 e^{2ik}}, \quad t = \frac{2J\gamma \sin k}{J^2 + \gamma^2 e^{2ik}}. \quad (10)$$

Figure 3(b) shows the schematic of the Hermitian conjugation counterpart $H_{c,2}^\dagger$ of the scattering system in Fig. 3(a). As $H_{c,2}^\dagger$ can be obtained by substituting $\gamma$ with $-\gamma$ in $H_{c,2}$, the scattering coefficients of $H_{c,2}^\dagger$ are given by Eq. (10) with $\gamma$ replaced by $-\gamma$. Notably, the reflection coefficients are identical for $H_{c,2}$ and $H_{c,2}^\dagger$, while the transmission coefficients are opposite.

The reflection and transmission of $H_{c,2}$ and $H_{c,2}^\dagger$ are plotted in Fig. 3(c), where the reflection is indicated by the red circles, the transmission is indicated by the blue squares. Notably, the total intensity is greater than unity, indicating that the scattering system extracts more energy from the environment via the scattering center than it loses to the environment. Furthermore, the difference between the reflection and transmission is always unity, which is verified by $|r|^2 - |t|^2 = 1$. In this case, $\bar{r}^*r$ and $\bar{t}^*t$ are real as plotted in Fig. 3(d). The numerical simulation of a Gaussian wave packet for the resonant input in $H_{c,2}$ is shown in Fig. 3(e), the reflection and transmission in the simulation are plotted in Fig. 3(f). As predicted by the energy-difference conservation relation, the reflection is always larger than the transmission by unity.

*Discussion.* Our findings are applicable to coupled microcavity systems that use fiber-taper waveguides as input and output channels. In the temporal coupled-mode theory [132,133], the scattering matrices for the two-port scattering center $H_c$ and its Hermitian conjugate $H_c^\dagger$ are

$$S(H_c) = \begin{pmatrix} s_{11} & s_{12} \\ s_{21} & s_{22} \end{pmatrix} = I - 2iD^\dagger \frac{1}{\omega - H_c + iDD^\dagger} D, \quad (11)$$

$$S(H_c^\dagger) = \begin{pmatrix} \bar{s}_{11} & \bar{s}_{12} \\ \bar{s}_{21} & \bar{s}_{22} \end{pmatrix} = I - 2iD^\dagger \frac{1}{\omega - H_c^\dagger + iDD^\dagger} D, \quad (12)$$

where $\omega$ is the input frequency, and $D_{lj}$ is a matrix of coupling coefficients between the microcavity mode $l$ and the asymptotic waveguide channel $j$ [34,35].

We still consider the scattering center $H_c$ to have the pseudo-Hermiticity expressed in Eq. (5), and the operator $q$ satisfies $q_{mj} = q_{jm} = \delta_{mj}$ and $q_{nj} = q_{jn} = \pm \delta_{jn}$. It is worth noting that $qDD^\dagger q^{-1} = DD^\dagger$, $qD = D\mathrm{diag}(q_{mm}, q_{nn})$, and $D^\dagger q^{-1} = \mathrm{diag}(q_{mm}^{-1}, q_{nn}^{-1})D^\dagger$. As a result, we obtain

$$S(H_c^\dagger) = \mathrm{diag}(q_{mm}, q_{nn})S(H_c)\mathrm{diag}(q_{mm}^{-1}, q_{nn}^{-1}), \quad (13)$$

which is a result of the specific pseudo-Hermiticity. This form of pseudo-Hermiticity ensures that the reflections and transmissions of $H_c$ and $H_c^\dagger$ are related by

$$\begin{pmatrix} \bar{s}_{11} & \bar{s}_{12} \\ \bar{s}_{21} & \bar{s}_{22} \end{pmatrix} = \begin{pmatrix} s_{11} & \pm s_{12} \\ \pm s_{21} & s_{22} \end{pmatrix}. \quad (14)$$

Substituting the reflections and transmissions into the conservation law $S^\dagger(H_c^\dagger)S(H_c) = I$ in Eq. (4), we obtain the energy conservation and the energy-difference conservation.

Finally, we emphasize that the conservation law is not limited to be valid for the two-port scattering system. We exemplify a three-port scattering system in the Supplemental Material [131]. Moreover, we can prove its validity for an arbitrary multiport scattering system from the perspective of the scattering matrix. Notably, the scattering matrix acting on the input yields the output. We have denoted $S(H_c)$ as the scattering matrix for the scattering center $H_c$. Then, the scattering matrix $[S^*(H_c)]^{-1}$ is the scattering matrix for the scattering center $H_c^*$, that is, $[S^*(H_c)]^{-1} = S(H_c^*)$. By taking the transpose of both sides of the above equation, we obtain $[S^\dagger(H_c)]^{-1} = S^T(H_c^*)$. Using the relation $S^T(H_c) = S(H_c^T)$ [85], we obtain $S^T(H_c^*) = S(H_c^\dagger)$. Thus, we have $[S^\dagger(H_c)]^{-1} = S(H_c^\dagger)$, which is equivalent to the conservation law $S^\dagger(H_c^\dagger)S(H_c) = I$.

*Conclusion.* We have found a conservation law expressed as $S^\dagger(H_c^\dagger)S(H_c) = I$ valid for any non-Hermitian scattering system. For two-port non-Hermitian scattering system, the conservation naturally appears as the reflections and transmissions of the non-Hermitian system $H_c$ and its Hermitian conjugation system $H_c^\dagger$ in the form of $\bar{r}^*r + \bar{t}^*t = 1$. The pseudo-Hermiticity of the non-Hermitian system links the sets $\{r, t\}$ and $\{\bar{r}, \bar{t}\}$. Consequently, the generalized conservation relation reduces to the energy-difference relation under the specific pseudo-Hermiticity protection. This explains the energy-difference behavior in the anti-$\mathcal{PT}$-symmetric scattering systems. To support our conclusion, we have demonstrated two prototypical anti-$\mathcal{PT}$-symmetric models, one with and one without pseudo-Hermiticity protection. Our findings provide profound insight on the pseudo-Hermiticity and stimulate further research interest in the anti-$\mathcal{PT}$-symmetric non-Hermitian systems, including the atomic gases, coupled resonators, electrical circuits, cavity magnonics, plasmonics, and beyond.

*Acknowledgment.* We acknowledge the support of the National Natural Science Foundation of China (Grant No. 12222504).